\documentclass[11pt]{article}

\usepackage{amsfonts,amsmath,amsxtra}
\usepackage{amssymb,amscd}
\usepackage[all]{xy}

\catcode`\@=11
\def\marginnote#1{}
\newcount\hour
\newcount\minute
\newtoks\amorpm
\hour=\time\divide\hour by60
\minute=\time{\multiply\hour by60 \global\advance\minute by-\hour}
\edef\standardtime{{\ifnum\hour<12 \global\amorpm={am}%
     \else\global\amorpm={pm}\advance\hour by-12 \fi
     \ifnum\hour=0 \hour=12 \fi
   \number\hour:\ifnum\minute<10 0\fi\number\minute\the\amorpm}}
\edef\militarytime{\number\hour:\ifnum\minute<10 0\fi\number\minute}
\def\draftlabel#1{{\@bsphack\if@filesw {\let\thepage\relax
\xdef\@gtempa{\write\@auxout{\string
   \newlabel{#1}{{\@currentlabel}{\thepage}}}}}\@gtempa
\if@nobreak \ifvmode\nobreak\fi\fi\fi\@esphack}
     \gdef\@eqnlabel{#1}}
\def\@eqnlabel{}
\def\@vacuum{}
\def\draftmarginnote#1{\marginpar{\raggedright\scriptsize\tt#1}}
\def\draft{\oddsidemargin -0.1truein
     \def\@oddfoot{\sl preliminary draft \hfil
     \rm\thepage\hfil\sl\today\quad\militarytime}
     \let\@evenfoot\@oddfoot \overfullrule 3pt
     \let\label=\draftlabel
     \let\marginnote=\draftmarginnote
\def\@eqnnum{{\rm (\theequation)}
\rlap{\kern\marginparsep\tt\@eqnlabel}%
\global\let\@eqnlabel\@vacuum}  }
\def\numberbysection{\@addtoreset{equation}{section}
     \def\theequation{\thesection.\arabic{equation}}}
\numberbysection

\renewcommand{\theequation}{\thesection.\arabic{equation}}
\makeatletter
\newdimen\normalarrayskip            
\newdimen\minarrayskip               
\normalarrayskip\baselineskip
\minarrayskip\jot
\newif\ifold             \oldtrue            \def\new{\oldfalse}
\def\arraymode{\ifold\relax\else\displaystyle\fi}
\def\eqnumphantom{\phantom{(\theequation)}} 
\def\@arrayskip{\ifold\baselineskip\z@\lineskip\z@
  \else
  \baselineskip\minarrayskip\lineskip1\baselineskip\fi}
\def\@arrayclassz{\ifcase \@lastchclass \@acolampacol \or
\@ampacol \or \or \or \@addamp \or
\@acolampacol \or \@firstampfalse \@acol \fi
\edef\@preamble{\@preamble
\ifcase \@chnum
  \hfil$\relax\arraymode\@sharp$\hfil
  \or $\relax\arraymode\@sharp$\hfil
  \or \hfil$\relax\arraymode\@sharp$\fi}}
\def\@array[#1]#2{\setbox\@arstrutbox=\hbox{\vrule
  height\arraystretch \ht\strutbox
  depth\arraystretch \dp\strutbox
width\z@}\@mkpream{#2}\edef\@preamble{\halign \noexpand\@halignto
\bgroup \tabskip\z@ \@arstrut \@preamble \tabskip\z@ \cr}%
\let\@startpbox\@@startpbox \let\@endpbox\@@endpbox
\if #1t\vtop \else \if#1b\vbox \else \vcenter \fi\fi
\bgroup \let\par\relax
\let\@sharp##\let\protect\relax
\@arrayskip\@preamble}
\def\eqnarray{\stepcounter{equation}%
           \let\@currentlabel=\theequation
           \global\@eqnswtrue
           \global\@eqcnt\z@
           \tabskip\@centering              
           \let\\=\@eqncr
           $$%
         \halign to \displaywidth  \bgroup
          \eqnumphantom \@eqnsel
   \hskip\@centering                               
 $\displaystyle  \tabskip\z@ {##}$%
 &\global\@eqcnt\@ne \hskip 2\arraycolsep
      $ \displaystyle  \arraymode{##}$\hfil
 &\global\@eqcnt\tw@ \hskip 2\arraycolsep
      $\displaystyle\tabskip\z@{##}$\hfil
      \tabskip\@centering
 &{##}\tabskip\z@\cr}
\makeatother

\newcounter{mo}

\newcounter{bk}

\newcommand{\Si}{\Sigma}
\newcommand{\tr}{{\rm tr}}
\newcommand{\ad}{{\rm ad}}
\newcommand{\Ad}{{\rm Ad}}
\newcommand{\ti}[1]{\tilde{#1}}
\newcommand{\om}{\omega}
\newcommand{\Om}{\Omega}
\newcommand{\de}{\delta}
\newcommand{\al}{\alpha}
\newcommand{\te}{\theta}
\newcommand{\vth}{\vartheta}
\newcommand{\be}{\beta}
\newcommand{\la}{\lambda}

\newcommand{\ve}{\varepsilon}
\newcommand{\ep}{\epsilon}
\newcommand{\vf}{\varphi}

\newcommand{\G}{\Gamma}
\newcommand{\ka}{\kappa}

\newcommand{\ga}{\gamma}

\newcommand{\si}{\sigma}

\newcommand{\mat}[4]{\left(\begin{array}{cc}{#1}&{#2}\\{#3}&{#4}
\end{array}\right)}

\def\bea{\begin{eqnarray}\new\begin{array}{cc}}
\def\ee{\end{array}\end{eqnarray}}

\newcommand{\beq}[1]{\begin{equation}\label{#1}}
\newcommand{\eq}{\end{equation}}
\newcommand{\beqn}[1]{\begin{small} \begin{eqnarray}\label{#1}}
\newcommand{\eqn}{\end{eqnarray} \end{small}}
\newcommand{\p}{\partial}
\def\sq2{\sqrt{2}}
\newcommand{\di}{{\rm diag}}
\newcommand{\oh}{\frac{1}{2}}

\def\sln{{\rm sl}(N, {\mathbb C})}
\def\sl2{{\rm sl}(2, {\mathbb C})}

\def\SLT{{\rm SL}(2, {\mathbb C})}

\def\f1#1{\frac{1}{#1}}

\newcommand{\bp}{\bar{\partial}}
\newcommand{\bz}{\bar{z}}

\newcommand{\bA}{\bar{A}}

\newcommand{\bL}{\bar{L}}

\def\mC{{\mathbb C}}
\def\mZ{{\mathbb Z}}
\def\mR{{\mathbb R}}

\def\frak{\mathfrak}

\def\gg{{\frak g}}

\def\gM{{\frak M}}

\def\gh{{\frak h}}

\def\bfe{{\bf e}}

\def\clA{\mathcal{A}}

\def\clC{\mathcal{C}}

\def\clG{\mathcal{G}}

\def\clR{\mathcal{R}}

\def\clO{\mathcal{O}}
\def\clH{\mathcal{H}}
\def\clK{\mathcal{K}}

\def\clL{\mathcal{L}}
\def\clM{\mathcal{M}}

\def\clP{\mathcal{P}}

\def\clS{\mathcal{S}}

\def\clX{\mathcal{X}}
\def\clY{\mathcal{Y}}
\def\clW{\mathcal{W}}

\def\bag2{{\bf g_2}}
\def\bas8{{\bf so(8)}}

\def\sr2{\sqrt{2}}
\newcommand{\ran}{\rangle}
\newcommand{\lan}{\langle}
\def\f1#1{\frac{1}{#1}}

\def\Ad{{\rm Ad}}

\def\ad{{\rm ad}}

\begin{document}


 \begin{flushright}
 ITEP-TH-04/22\\
 IITP-TH-03/22
 \end{flushright}
\vspace{3mm}

 \begin{center}
{\Large\bf 2d Integrable systems, 4d Chern-Simons theory}
\\ \vspace{4mm}
{\Large\bf and Affine Higgs bundles}\\
 \vspace{10mm}
  {{\Large A. Levin}$^{\,\natural\,\,\flat}$\ \
  {\Large M. Olshanetsky}$^{\,\flat\,\S}$\ \  {\Large A. Zotov}$^{\,\diamondsuit\,\flat\, \natural}$}\\
   \vspace{8mm}
  \vspace{2mm}$^\flat$ -
  {\rm Institute of Theoretical and Experimental Physics  NRCKI,\\ B. Cheremushkinskaya, 25, Moscow, 117259, Russia}\\
   \vspace{2mm}$^\S$ - {\rm Institute for Information Transmission Problems RAS (Kharkevich Institute),
 \\  Bolshoy Karetny per. 19, Moscow, 127994,  Russia}\\
 \vspace{2mm}$^\natural$ - {\rm National Research University Higher School of Economics, Russian Federation,\\
 Usacheva str. 6,  Moscow, 119048, Russia}\\
  \vspace{2mm} $^\diamondsuit$ -
 {\rm Steklov Mathematical Institute of Russian
Academy of Sciences,\\ Gubkina str. 8, Moscow, 119991,  Russia
 }\\

 \vspace{4mm}
 {\footnotesize Emails: alevin2@hse.ru, olshanet@itep.ru,
 zotov@mi-ras.ru}
 \end{center}

 \begin{abstract}
We compare constructions of
2d integrable models through two gauge field theories. The first one is the 4d Chern-Simons (4d-CS) theory proposed by Costello and Yamazaki.
The  second one is the 2d generalization of the Hitchin integrable systems constructed by means
 the Affine Higgs bundles (AHB). We illustrate the latter approach by considering 1+1 field versions of integrable systems including
 the Calogero-Moser field theory, the Landau-Lifshitz model and the field theory generalization of the elliptic Gaudin
 model.
 \end{abstract}

\footnotesize \tableofcontents \normalsize

\section{Introduction}

In the nineties, we attempted to construct 2D classical integrable field theories starting with a two-dimensional WZW action \cite{GOS}.
The corresponding equations of motion coincide with the Zakharov-Shabat equations.
 These equations are the hallmark of two-dimensional integrable systems. But that approach had
 one essential drawback --
the Lax operator did not depend on the spectral parameter. This parameter is a necessary ingredient for constructing the infinite number of commuting integrals of motion.
A class of integrable theories, derived from the WZW models was considered in papers by L. Feh\'{e}r
et all (see the review \cite{F}). Also, the interrelations between gauge theories and integrable systems were
considered in the mid-nineties in \cite{N1,N2}. Later Nekrasov and Shatashvili derived quantum integrable
systems from four-dimensional gauge theories \cite{N3}.

The problem with the spectral parameter was overcome in the works of Costello and Yamazaki  \cite{CY} by considering the so-called four-dimensional Chern-Simons theory (4d-CS).

Here we compare 4d-CS construction with the construction of 2d integrable systems based on the Affine Higgs bundles (AHB) model proposed in \cite{LOZ}. The AHB model is the 2d analog of the Hitchin
systems \cite{Hi}.
To compare the AHB theory with the 4d-CS approach
 we rewrite the AHB theory in the form of a special 4d CS model. It allows one to establish
 a correspondence between the field contents from  both constructions.

The first formal difference between these two approaches is that AHB theory is free, and the nontrivial integrable models appear as a
result of the symplectic reduction. The latter procedure is similar to  what happens in the finite-dimensional case for the Hitchin systems. Symplectic reduction is defined by two types of constraints. The first one is given by the moment map constraints (the Gauss law analog in the YM theory).
The second one is the gauge fixing conditions.
After imposing these constraints we come to the symplectic phase spaces of 2d integrable systems.
Using the AHB we constructed in \cite{LOZ} the 2d field generalization of the elliptic (spin) Calogero-Moser (CM) model.
It was proved by A. Shabat (unpublished) and in \cite{AZ} that this model is gauge equivalent
to the Landau-Lifshitz (LL) equation \cite{Skl}. The gauge transformation comes from the so-called symplectic Hecke correspondence.
Another example of 2d generalization of the Hitchin systems is 2d elliptic Gaudin model. In particular, the Principal Chiral Model is reproduced in this way
\footnote{In a recent paper \cite{VW} authors proposed an approach to the affine Gaudin models based on the 3d BF theory that is very close to the AHB construction.}.

Another construction similar to the AHB approach is the algebra-geometric derivation of the Zakharov-Shabat equation proposed by Krichever
\cite{Kr}. In particular, using the KP hierarchy he constructed the 2d version of the  Calogero-Moser model. This approach can be also extended to the field version of the Ruijsenaars-Schneider models \cite{ZZ}.

In contrast to AHB construction, the 4d-CS theory is not free. The equations of motion
have the form of the moment map constraints equations, which are similar to the moment map constraints
in the AHB theory. It only remains to impose some gauge fixation to
come to  2d integrable systems.
To compare these constructions, we rewrite the equations of motion and the moment map constraints in the AHB models in the CS form.

In the  standard approach to the 2d integrable in \cite{CY,LOZ} the 3d space has the form $\mR\times\mC P^1$ or
$S^1\times\mC P^1$ or with an elliptic curve instead of $\mC P^1$. More generally, these 3d spaces can be replaced
by an arbitrary Seifert surface \cite{Se}. The  Seifert surface is a $U(1)$ bundle over the Riemann curve
 $\Si_g$ of genus $g$.
The Seifert surfaces have two topological characteristics $(n,g)$, where $n$ is the degree of the line bundle
corresponding to the  $U(1)$ bundle. Although the moduli space of the Higgs bundles over
 the Seifert surfaces depends on $n$, the invariant Hamiltonians do not depend on it. The reason is that there
 exists singular gauge transformation $\Xi(k)$ of the Lax operator $L(n)$ such that $\Xi(k)\,:\,L(n)\to L(n+k)$.

The AHB construction allows one to define 2d analogs of the  additional structures in the
Hitchin systems. The first structure  is the affine analog of the symplectic Hecke correspondence \cite{AZ,LOZ}.
Another structure that appears in the AHB model is the affine version of the Nahm equations describing the surface defects.
Both of these structures
 will be considered in the forthcoming publication \cite{LOZ22}.

The paper is organized as follows. In the next section we explain briefly 4d-CS construction of 2d integrable models based on the articles \cite{CY,L}. In Section 3  the AHB construction is given following
 notations from \cite{LOZ,Z}. Some examples are given in Section 4.
  Finally, we establish the correspondence between the two construction in Section 5.


\section{4d Chern-Simons model and integrable systems}
Let us describe the field content of 4d Chern-Simons model. Consider a Riemann curve $C$ and the space time
$M=\mR^2\times C$ with the local coordinates $(x,t)$, $(z.\bz)$
\footnote{Here we follow notations from \cite{CY}.}. On $\mR^2\sim\mC$ introduce the
 complex coordinates $w=x+t$, $\bar w=x-t$.
Let $G$ be a complex simple Lie group. Consider a principal $G$ bundle $\clP$ over $M$ and
equip it with the connections
\beq{coa}
d+A=(\p_w+ A_w)\otimes dw+(\p_{\bar w}+A_{\bar w})\otimes d{\bar w}+
(\bp+\bA)\otimes d\bz=A_tdt+A_xdx+A_{\bz}d\bz\,.
\eq
Let  $\om$ be a one form on $\Si$ ($\om=\varphi(z)dz$). It is a section of the canonical class
$\clK_C$  on $C$.
The four-dimensional CS action is defined as
\beq{cs4}
S_{4d} =\f1{2\pi\hbar}\int_M\om\wedge CS(A)\,,
\eq
where $CS(A)$ is the standard CS action
$$
CS(A) := \tr\Bigl( A \wedge dA +\frac{2}{3} A \wedge  A \wedge  A\Bigr)
$$
and $A$ is the defined above connection (\ref{coa}).

Beyond the points where the form $\om$ vanishes the
equations of motion corresponding to (\ref{cs4}) take the form:
\beq{ecs}
\begin{array}{ll}
        1.&  [D_{A_w},D_{A_{\bar w}}]=0\,, \\
2.& [D_{A_w},D_{A_{\bz}}]=0\,, \\
3. & [D_{A_{\bar w}},D_{A_{\bz}}]=0 \,.
        \end{array}
\eq
These
equations are invariant under the gauge transformations
\beq{ga}
A\to A^f=f(d+A)f^{-1}\,,
\eq
\beq{gg}
f\in\clG=C^\infty (M\to G)\,.
\eq

Let $f$ be the gauge transformation fixing the gauge as $A_{\bz}^f=A_{\bz}^0$. We
identify $A^f_w=L(w,\bar w,z)$ with the Lax operator, and $A^f_{\bar w}=M( w,\bar w,z)$ with the
 evolution operator $M$.  Then the first equation in (\ref{ecs}) turns into  the Zakharov-Shabat type equation for some 2d integrable system:
\beq{zs}
\p_{\bar w}L-\p_w M+[M,L]=0\,.
\eq

In the most part of the paper \cite{CY} it is assumed that there is a gauge choice
\beq{cg}
A_{\bz}=0\,,
\eq
or, put it differently, that the moduli space of holomorphic bundles over $C$ is empty.
It is indeed true if $C$ is a rational curve, but almost never true in the general case.
For example, if $C$ is an elliptic curve this is possible for the topologically non-trivial bundles.
If it is the case, then
the equations 2 and 3 from (\ref{ecs}) mean that
 $A_w$ and $A_{\bar w}$ are holomorphic on  $C$ and in this way they are constants.
 Therefore, we are left with the  Zakharov-Shabat equation,
 where the operators $L$ and $M$  are independent  on the spectral parameter $z$.

In order to come to meaningful cases with $L$ and $M$ depending on the spectral parameter one should consider higher genus curves. One more possibility is to consider additional degrees
of freedom by introducing surface defects in the 4d-CS model. The surface defects come from the poles
and zeros of the meromorphic 1-form $\om$ in (\ref{cs4}).
The zeros of $\om$ mean that the Lax operator has poles at this points and the corresponding
coefficients (residues) define additional degrees of freedom in the theory. These defects are called the
disorder defects.

The poles of $\om$ lead to restrictions of the gauge fields
at these poles and also add degrees of freedom.
These defects are called the order defects.
Below we consider these defects in terms of AHB theory in greater detail.
%
%
%


\section {Affine Higgs bundle}

\subsection{Three-dimensional space}

Consider a principal  $U(1)$-bundle $W$ over Riemann curve $\Si$:
\beq{wb}
W\stackrel{\pi}\rightarrow\Si\,,~~(W=U(1)\to\Si)\,.
\eq
The total space of the bundle is called the \emph{Seifert surface}.
  Let $(z,\bz,\te)$ be local coordinates on $W$
     and $\Om^{(m,n,k)}(W)$ the space of corresponding $(m,n,k)$-forms.
     Redefine the one forms as
  \beq{di}
  d\ti\bz= d\bz\,,\quad d\ti\te=d\te-n\bar\mu(z,\bz)d\bz\,.
  \eq
Here $n$ is the degree of the $U_1$-bundle and $\bar\mu(z,\bz)\in \Om^{(0,-1,1)}$ is the Beltrami
differential.
Consider    $\Om^{(1,0)}(\Si)$-form $dz$ on $\Si$ and let $\pi^*(dz)\in\Om^{(1,0,0)}(W)$.
 Define two vector fields on $W$, which annihilate the form $\pi^*dz$:
 $$
 1.\, \p_\te\,,\quad
  2.\,\p^{\bar\mu}_{\bz}\,.
  $$
The first field $\p_\te$ acts along the $S^1$ fibers and thereby annihilates the form $\pi^*dz$.
For the second field $\p^{\bar\mu}_{\bz}$ this condition
 means that
\beq{pm}
\p^{\bar\mu}_{\bz}=\p_{\bz}+n\bar\mu(z,\bz)\p_\te\,.
\eq
Let
\beq{tt}
\ti\te=\te-n\int^{\bz}\bar\mu(z,\bz)
\eq
be a local coordinate in the bundle $W$. Then for a smooth function $f$
\beq{at}
\p^{\bar\mu}_{\bz}f(\ti\te)=0\,.
\eq

Consider a line bundle $\clL$ over $\Si_g$, which is a complexification of the $U(1)$-bundle.
Let $D_z\subset\Si_g$ be a small disc with the center $z=0$ and $D'_z\subset D_z$
The degree $n$ of the bundle is defined by a holomorphic non-vanishing transition function
$f(z)$ on $D_z\backslash D'_z$.
The degree can be changed by the multiplication $f(z)\to f(z)w(z)$ in the following way.
\beq{mod}
\te\to\te- k\cdot{\rm arg}(w)\,.
\eq
This procedure is called \emph{the modification} of the $U(1)$-bundle.

If the bundle $W$ is trivial then  one can take
$n=0$. In the examples below we assume  $n=0$.

Let $G$ be a complex Lie group and $\clP$ is a principle $G$-bundle over $W$. We define preliminary,
 \emph{the affine Higgs bundle} (AHB) over $W$ as  a pair of connections
\beq{ahb}
( D_{\bA,\bar \mu}=\p^{\bar\mu}_{\bz}+A_{\bz}\,,\,\p_\te+ A_\te)\,.
\eq
The first component $\p^{\bar\mu}_{\bz}+A_{\bz}$ defines the complex structure on the sections
of $\clP$ in $(\bz,\te)$ direction.
The precise definition of the AHB is given below (\ref{cfc}).
The second component  is \emph{the Higgs connection}. It is an affine analogue of the Higgs field introduced
by Hitchin \cite{Hi2}.

\subsection{Affine holomorphic bundles}

The affine Higgs bundles are the cotangent bundles to the affine holomorphic bundles,
which we are going to define.

%
In the previous subsection we introduced the connection acting on the sections $\G(\clP)$ (\ref{ahb}):
$$
  D_{\bA,\bar \mu}=(\p_{\bz}+\bar \mu(z,\bz)\p+ \bA(z,\bz,x))\otimes d\bz\,.
$$
%
%
Consider, in addition,  a line bundle $\clL$ over $\Si$ with the connection $(\p_{\bz}+\bar k_{\bz})\otimes d\bz$.
 The anti-holomorphic connection on $\clP\oplus\clL$ is the pair of
 operators
 \beq{ac}
\nabla_{\bA,\bar \mu,\bar k}=\left(
\begin{array}{c}
 D_{\bA,\bar \mu} \\
 (\p_{\bz}+ \bar k(z,\bz))\otimes d\bz
\end{array}
\right)\,.
 \eq
Let  $G(W)$ be a smooth map of $W$ to $G $
$$
G(W)=C^\infty(W\to G )\,,~~G (W)=\left\{\sum_jf_j(z,\bz)e^{j\te},\,|\,f_j\in C^\infty(\Si\to G)\right\}\,.
$$
It can be considered as a map of the spectral curve $\Si$ to the loop group
\beq{stg}
G (W)=C^\infty(\Si\to  L(G ))\,.
\eq
 The structure group of the bundle $\clP \oplus\clL$ (the gauge group) is defined by
 replacing $L(G )$ with its central and co-central extensions (\ref{cce}):
 $$
 \check {\cal G}:=C^\infty(\Si\to  \check L(G ))\,.
 $$
 More precisely,
 \beq{gg1}
\check {\cal G}=
(G (W),\{\exp(\varepsilon_3(z,\bz)\})\rtimes\{\exp(\varepsilon_2(z,\bz)\p_\te\}\,,
\eq
$$
\varepsilon_{2}(z,\bz)\p_\te\in C^\infty(\Si\to\mC)\,,
~\varepsilon_{3}(z,\bz)\in C^\infty(\Si\to\mC)\,.
$$
%
Consider its infinitesimal action on  $\nabla_{\bA,\bar\mu,\bar k}$.
As a vector space the Lie algebra Lie$(\hat {\cal G}^{G })$
has three components:
\beq{lg}
Lie(\hat {\cal G})=M_1\oplus M_2\oplus M_3\,,
\eq
$$
M_1=C^\infty (M \to \gg)=\{\ep_1(z,\bz,\te)\}\,,
$$
$$
M_2=C^\infty (\Si\to \mC) =
\{\varepsilon_2(z,\bz)\p_\te \}\,,
$$
$$
 M_3= C^\infty (\Si \to \mC) )=\{\varepsilon_3(z,\bz)\})\,.
 $$
 Their action on $\nabla_{\bA}$ takes the form:
 \beq{hvf}
\begin{array}{llll}
1.& \de_{\ep_1} \bA= - (\p_{\bz}+\bar\mu\p)\ep_1+[\ep_1,\bA]\,, &\de_{\ep_1}\bar\mu=0 \,,& \de_{\ep_1} \bar k=\lan \bA\p \ep_1\ran \,,\\
2.& \de_{\varepsilon_{2}} \bA=\varepsilon_{2}\p_\te \bA\,,&   \de_{\varepsilon_2} \bar\mu=
 -\p_{\bz}\varepsilon_{2}\,,  &\de_{\varepsilon_2} \bar k=0\,,\\
3.& \de_{\varepsilon_{3}}\bA=0\,,&\de_{\varepsilon_3} \bar\mu=0\,,  &
 \de_{\varepsilon_3} \bar k=-\p_{\bz}\varepsilon_3
\,.
 \end{array}
 \eq
%
 The moduli of holomorphic structure on $\clP(M)\oplus\cal L$ is the quotient space
 \beq{bun1}
 Bun_{G ,M}=\nabla_{\bA,\bar \mu,\bar k}/\check {\cal G}=
 \nabla_{\bL,\bar \mu,\bar k}\,,
 \eq
where we
  fix the gauge as $\bA\to \bA^f=\bL$, i.e.
 \beq{fg}
 \bL =\bA^f=f\p_{\bz} f^{-1}+f\bA f^{-1}\,.
 \eq
 One can fix the action of the abelian subgroups
 $\{\exp(\varepsilon_3)\}$, $\{\exp(\varepsilon_3(z,\bz)\p)\}$ on $\bar\mu$ and $\bar k$
 (\ref{hvf}) in a similar way.
 We preserve the notations for the gauge transformed variables $\bar \mu$ and $\bar k$.



\subsubsection{Affine Higgs bundles}

Introduce the Higgs field $\Phi(z,\bz,\te)$.
Let $\clK$ be a canonical class of $\Si$. Then the Higgs field  is
$\Phi(z,\bz,\te)\in C^\infty(\Si\to (L(\gg)\otimes d\te)\otimes\clK$.


  Let $\nu(z,\bz),\,r(z,\bz)\in\Om^{(1,0)}(\Si)$
Define
  \beq{hf1}
 \nabla_{\Phi,\nu,r} =\left(
              \begin{array}{c}
             D_{\Phi,\nu}   \\
                r(z,\bz) \\
              \end{array}
           \right)\otimes \clK\,,
 \eq
 \beq{nu}
 D_{\Phi,\nu}= (\nu(z,\bz)\p_\te +\Phi(z,\bz,x))d\te\,.
 \eq
 The affine Higgs bundle is the pair
\beq{cfc}
\clH^{aff}(G )=(\nabla_{\bA,\bar\mu,\bar k}\,,\,\nabla_{\Phi,\nu,r})\sim T^*\nabla_{\bA,\bar\mu,\bar k}=\{\bA\,,\,\bar\mu\,,\,\bar k\,,
\,\Phi\,,\,\nu\,,\,r\} \,.
\eq
The connection form $A_\te$ in (\ref{ahb}) is related to the Higgs field $\Phi$ as
\beq{ap}
A_\te=\frac{\Phi}\nu\,.
\eq
 The fields of the Higgs bundles
have the following dimensions:
\begin{center}
\textbf{Table 1}: Dimensions of fields
\end{center}
\vspace{-0.5cm}
$$
\begin{tabular}{|c |c|c|c|}
  \hline
    &$z$& $\bz$&$\te$\\
    \hline
    \hline
  $\bA$ & 0 & 1 & 0 \\
  $A_\te$ & 0 & 0 & 1 \\
  $\bar\mu$ & 0 & 1 & -1 \\
  $\bar k$ & 0 & 1 & 0 \\
  $\Phi$ & 1 & 0 & 1 \\
  $\nu$ & 1 & 0 & 0 \\
  $r$ & 1 & 0 & 1 \\
  \hline
\end{tabular}
$$

The cotangent bundle structure of the AHB comes from the pairing (\ref{pa})
$\clH^{aff}(G )=T^*\nabla_{\bA,\bar\mu,\bar k}$.

\noindent Define the symplectic form $\Om$ on $\clH^{aff}(G )$
\beq{bomi1}
\Om=\frac{1}{\pi}\int_\Si|d^2z|\Bigl(\lan\de\Phi,\de \bA\ran
+\de r\de\bar\mu+\de\nu\de\bar k\Bigr)\,,
\eq
where
$$
\lan\de\Phi,\de \bA\ran=\f1{2\pi}\int_{S^1}(\de\Phi,\de \bA)\,.
$$
The form is invariant under the action of the gauge group  $\hat {\cal G}$ (\ref{gg1}).
 Along with (\ref{hvf}), the corresponding Hamiltonian vector fields are as follows:
\beq{hvf1}
\begin{array}{llll}
1.\,& \de_{\ep_1} \Phi=  \nu\p_\te \ep_1+[\Phi,\ep_1]\,, & \de_{\ep_1}\nu=0\,,
&\de_{\ep_1} r=\lan \Phi,\p\ep_1\ran\,, \\
    2.\,&\de_{\varepsilon_{2}} \Phi= \varepsilon_{2} \p_\te\Phi\,,&
\de_{\varepsilon_2} \nu=0\,, &\de_{\varepsilon_2} r=0\,,\\
   3.\,& \de_{\varepsilon_3}\Phi=0\,,  &
 \de_{\varepsilon_3} \nu=0\,, &\de_{\varepsilon_3} r=0\,.\\
   \end{array}
    \eq
%
%
%
%
The action of $\hat {\cal G}$ is generated by the moment maps
$m_j\,:\,\clH(G )\to Lie^*(\hat {\cal G})$, where
\beq{ld}
Lie^*(\hat {\cal G}^{G })=M^*_1\oplus (M^*_2\sim M_3)\oplus (M^*_3\sim M_2)\,.
\eq
More explicitly,
$$
\begin{array}{l}
m_1=(\p_{\bz}+\bar\mu\p_\te))\Phi- \nu\p_\te \bA+[\bA,\Phi]\in M_1^*\,,\\
(m_1=[D_{\bA,\bar\mu,\bar k},D_{\Phi,\nu,r}])\,,\\
m_2=\int_{S^1}\lan\p_\te\Phi,\bA\ran- \p_{\bz}r\in M_2^*\,,\\
m_3=\p_{\bz}\nu\in M_3^*\,.
\end{array}
$$
Let $\clC^{aff}(\bA,\bar\mu,\bar k|\Phi,\nu,r)$ be the  set of solutions
 of the moment equations $m_j=0$, $(j=1,2,3)$
\beq{me1}
\left\{
\begin{array}{l}
(\p_{\bz}+\bar\mu\p_\te)\Phi- \nu\p_\te \bA+[\bA,\Phi]=0\,,\\
([D_{\bA,\bar\mu,\bar k},D_{\Phi,\nu,r}]=0)\,,\\
m_2=\int_{S^1}\lan\p_\te\Phi,\bA\ran- \p_{\bz}r=0\,,\\
m_3=\p_{\bz}\nu=0\,.
\end{array}
\right.
\eq
The quotient of $\clC^{aff}$ under the action of the gauge group $\check {\cal G}$ (\ref{gg1}) is the
moduli space of  the affine Higgs bundles:
\beq{ms1}
\gM^{aff}(G )=\clH^{aff}(G )//\check {\cal G}\sim\clC^{aff}/\check {\cal G}\,.
\eq
%
%
%
 We can first fix the gauge and then solve
the moment map equations. In this respect
$\gM^{aff}(G )$ is  defined as the set of solutions of equations
\beq{clm}
(\p_{\bz}+\bar\mu\p_\te)L- \nu\p_\te \bL+[\bL,L]=0\,,
\eq
$$
\p_{\bz}r=\int_{S^1}\lan\p_\te L,\bL\ran\,,\quad \p_{\bz}\nu=0\,.
$$


\subsubsection{Parabolic structures. The order defects.}


 To introduce the parabolic structure we
  attach  the coadjoint orbits
  ${\mathcal O}_a={\mathcal O}(p_a^{(0)},c_a^{(0)})$ of the loop group $L(G )$ (\ref{c11})
  to the marked points $z_a\in\Si$, $a=1,\ldots,n$.
  It means that we add the order defects in the theory.
  The disorder defects correspond to the
  reducing the gauge group
$\check {\cal G}$ (\ref{gg1}) to the subgroup
$\check {\cal G}(\times_a Fl_a)\subset \check {\cal G}$,
which  preserves the affine flags $Fl_a$ at the marked points. It was proved in \cite{LOZ2}
that these construction are equivalent. Here we follow the order defects
description.

The affine parabolic Higgs bundle has the following field contents:
\beq{cfc1}
\clH^{aff,\,par}(G )=(\bA\,,\,\bar\mu\,,\,\bar k\,,\,\Phi\,,\,\nu\,,\,r\,,\,\cup_{a=1}^n \clO_a) \,.
\eq
The coadjoint orbits (\ref{c11}) are equipped with the Kirillov-Kostant  symplectic form (\ref{1.6a}).
Thereby, the symplectic form on the reduced parabolic Higgs bundle $\clH^{aff\,par}(G )$  is
equal to
 \beq{psf}
\Om-\sum_{a=1}^n\om_a(p_a^{(0)},c_a^{(0)})\,,
\eq
where $\Om$ is the form (\ref{bomi1}) and $\om_a$ are the Kirillov-Kostant forms (\ref{1.6a}).
Due to the presence of new terms in the form, the moment map constraints (\ref{me1})
 are upgraded as
 $$
 m_1=\sum_{a=1}^nS(p_a^{(0)},c_a^{(0)})\de(z-z_a,\bz-\bz_a)\,,~~
 m_3=\sum_{a=1}^nc_a^{(0)}\de(z-z_a,\bz-\bz_a)\,,
 $$
 so that
\beq{co11}
\p_{\bz}\Phi- \nu\p_\te A_{\bz}+[\bA,\Phi]=\sum_{a=1}^nS(p_a^{(0)},c_a^{(0)})\de(z-z_a,\bz-\bz_a)\,,
\eq
\beq{co12}
\p_{\bz}\nu=\sum_{a=1}^nc_a^{(0)}\de(z-z_a,\bz-\bz_a)\,.
\eq
It means that $\nu$ is not a constant in (\ref{co11}) but a meromorphic $(1,0)$-form on $\Si$ with the first order poles at $z=z_a$:
\beq{ln}
\nu|_{z\to z_a}\sim\frac{c_a^0}{z-z_a}\,.
\eq
In other words, $\nu=const$ implies that we deal with orbits without central extension only, i.e.
\beq{to}
S_a=gp_a^{(0)}g^{-1}\,.
\eq
Since $\sum_{a=1}^nc_a^{(0)}=0$, in the case of a single marked point (likewise it happens for
the Landau-Lifshitz equation) the orbit has the form (\ref{to}) and $\nu=\nu^0$ is a constant.

Next, we pass to the symplectic quotient (the moduli space).
Let us fix a gauge as in (\ref{fg}) and
\beq{fg1}
L=\nu f^{-1}\p_\te f+f^{-1}\Phi f\,,~~(f\in\hat {\cal G})\,,
\eq
\beq{fg2}
L/\nu=f^{-1}\p_\te f+f^{-1}A_\te f\,.
\eq
The moment map constraint equation (\ref{co11})
with $m_1=0$ is modified as
\beq{hb2}
\p_{\bz}L-\nu\p_\te \bar L+[\bar L,L]=\sum_{a=1}^n\de(z-z_a)S_a\,,~~
\left( [D_{\bL,\bar\mu},D_{L,\nu}]=\sum_{a=1}^n\de(z-z_a)S_a\right)\,.
\eq
%
Solutions of this equation along with (\ref{co12}) define the moduli space of the affine parabolic bundles as  the symplectic quotient space
\beq{ms}
\clH^{aff,\,par}(G )//{\mathcal G}\sim \gM^{aff,\,par}(G )\,.
\eq
It is a phase space of 2d integrable systems.
The symplectic form (\ref{psf}) on $\clM^{aff,\,par}(G )$ turns into (see (\ref{psf}))
\beq{1.4a}
\Omega^{par}=\int_{\Sigma}\left(\lan\delta L|\delta\bL\ran+
\delta\nu\delta \bar k+\de r\de\bar\mu\right)
-\sum_{\al=1}^n\omega_\al\,.
\eq


\subsection{Equations of motion}

 Let $W=S^1\times\Si$ be a trivial bundle. The measure on $W$ is $\varpi(z,\bz)d\te$,
where  $\varpi(z,\bz)\in\Om^{(1,1)}(\Si)$ is a $(1,1)$-form on $\Si$.
The gauge invariant integrals are generated by the traces of the monodromies of the Higgs field $A_\te$. We take the Hamiltonian in the form:
\beq{1.15}
H(\Phi,\nu)=\int_\Si\varpi(z,\bz) \left(\tr\exp\oint_{S^1}  A_\te(z,\bz,\te)\right)=
\eq
$$
\int_\Si\varpi(z,\bz) \left(\tr\exp\f1{\nu(z,\bz)}\oint_{S^1}  \Phi(z,\bz,\te)\right)\,.
$$
%
%
%
Consider equations of motion on the "upstairs" space $\clH^{aff}(G )$
(\ref{cfc1}).
They are derived by means of the symplectic form  (\ref{psf})
 and the Hamiltonians (\ref{1.15}).
 In this way we obtain the following  free system:
\beq{1.20}
\dot{\Phi}=0\,,
\eq
\beq{1.21}
\dot{\bA}(z,\bz)=
\frac{\de \clH}{\de \Phi(z,\bz,\te)}=\frac{\varpi(z,\bz)}{\nu(z,\bz)}\exp\f1{\nu(z,\bz)}\oint_{S^1}\Phi(z,\bz,\te)d\te\,,
\eq
\beq{1.22}
\dot{\nu}=0\,,~~\dot{\bar\mu}=0\,,~~\dot{r}=0\,,
\eq
\beq{1.22a}
\dot{\bar k}(z,\bz)=\frac{\de \clH}{\de \nu(z,\bz)}=-\frac{\varpi(z,\bz)}{\nu^2(z,\bz)}\oint_{S^1} \Phi(z,\bz,\te))d\te
\exp\int_{S^1} \frac{\Phi(z,\bz,\te)}{\nu(z,\bz)}d\te\,.
\eq

Recall that
after the symplectic reduction we come to the fields $\bL$ (\ref{fg})
and $L$ (\ref{fg1}). For simplicity,
we keep the same notation for the coadjoint orbits variables $S_\al$,
so they are transformed as in (\ref{kki}).
 This yields
\beq{1.17}
H(L,\nu)=\int_\Si\om(z,\bz)  \left(\tr\exp\f1{\nu(z,\bz)}\oint_{S^1} d\te L(z,\bz,\te)\right)\,.
\eq

 Let $W$ be a non-trivial bundle  $(n\neq 0)$.
 It follows from (\ref{at}) that  $\bL$ depends on $\ti\te$ (\ref{tt}).
 The moment equation (\ref{hb2}) takes the  form
 $$
 (\p_{\bz}+n\bar\mu\p_\te)L-\nu\p_\te \bar L+[\bar L,L]=\sum_{a=1}^n\de(z-z_a)S_a\,,~~
\left( [D_{\bL,\bar\mu},D_{L,\nu}]=\sum_{a=1}^n\de(z-z_a)S_a\right)\,,
 $$
Its solution $L$ has the same form  as for $n=0$, but the angle parameter $\te$
is replaced with $\ti\te$.
 The corresponding monodromy matrix  is conjugated to the original monodromy matrix
 $$
 \exp\f1{\nu(z,\bz)}\oint_{S^1} d\te L(z,\bz,\ti\te)=\Xi(n)\left(\exp\f1{\nu(z,\bz)}\oint_{S^1} d\te L(z,\bz,\te)
 \right)\Xi(n)^{-1}\,,
 $$
 where the gauge transformation assumes the form
 $$
 \Xi(n)=\exp\int_0^\de  d\te L(z,\bz,\te)\,,~~\de= n\int^{\bz}\bar\mu\,.
 $$
In this way, as we claimed in the Introduction, the invariants of the monodromy matrix and, in particular, the Hamiltonian are independent of $n$.


It follows from the moment map equation (\ref{hb2}) that for the parabolic bundles the Lax operator $L$ has the first order poles
at the marked points $z_a$.
 Let $w_a=z-z_a$.
 The generating function of the Hamiltonians (\ref{1.17}) has the expansion:
\beq{1.16}
H(L,\nu)=\sum_{a\in I}\sum_{j=-1}^{+\infty}H^a_jw_a^j\,.
\eq




 Consider the set of times $T_{a,j}=\{t_{a,j}\}$ corresponding to the Hamiltonians $H^a_j$.  The one-dimensional spaces $T_{a,j}$ are isomorphic to $\mR$.
Let $\p_{a,j}=\{H^a_j,~\}$ be the Poisson vector field on the moduli space
$\gM^{aff,\,par}(G )$ (\ref{ms}).
Assume that the gauge transformation $f$ comes from the gauge fixation (\ref{fg}).
Define the connection form $M_{a,j}=\p_{a,j}ff^{-1}$.
  From (\ref{fg1}) we have
$\Phi=-\nu\p f f^{-1} + fLf^{-1}$.
Plugging it into (\ref{1.20}) we come to the Zakharov-Shabat
equation
\beq{1.23}
  \p_{a,j}L-\nu\p_\te M_{a,j}+[M_{a,j},L]=0\,,~~\left([D_{M_{a,j}},D_{L}] =0\right)\,,
\eq
where $D_{M_{a,j}}=\p_{a,j}+M_{a,j}$.
Notice that
 the variables on the moduli space $L,\bL,S_a$ do not depend on $\bar k$. In this way
 the dynamics of $\bar k$ (\ref{1.22a}) is inessential.
 The operators $M_{a,j}$ can be restored partly from the
 equation (\ref{1.21}):
\beq{1.24}
\bp M_{a,j}-\p_{a,j}\bL+[M_{a,j},\bL]=\frac{\de H^a_j}{\de L}\,,~~
\left([D_{\bL},D_{M_{a,j}}] =\frac{\de H^a_j}{\de L}\right)\,,
\eq
where
\beq{lf}
\frac{\de H^a_j(L)}{\de L}=f\frac{\de H^a_j(\Phi)}{\de \Phi}f^{-1}
\eq

The   equations (\ref{1.23}) and (\ref{1.24}) along with the moment constraint equation (\ref{hb2}) yield the system:
\beq{cc}
\begin{array}{ll}
        1.& [D_{M_{a,j}},D_{L,\nu}] =0\,,  \\
2.& \displaystyle{
[D_{\bL},D_{M_{a,j}}] =\frac{\de H^a_j}{\de L}
}
\,, \\
3. &  [D_{\bL},D_{L,\nu}]=\sum\limits_{a=1}^n\de(z-z_a)S_a \,.
        \end{array}
\eq
%
Let $V$ be a module of the Lie algebra $\gg$.
 Consider the associated bundle $E=\clP\times_GV$, where $\clP$ is the principle $G$-bundle over $W$. Equivalently, we can consider the associated vector $L(G)$-bundle
over $\Si$. Let $\Psi$ be a section of $E$.
Consider the linear system
\beq{ls1}
\begin{array}{ll}
        1.&(\nu\p+L)\Psi  =0\,, \\
2.& (\p_{\bz}+\bL)\Psi=0\,, \\
3. &  (\p_{a,j}+M_{a,j})\Psi=0\,.
        \end{array}
\eq
Then the equation (1.\ref{cc}) is the consistency condition for the equations 1. and 3.
and the equation (3.\ref{cc}) is the consistency conditions for the equations 1. and 2.\,.

\subsection{Conservation laws}

The matrix equation (1.\ref{ls1}) allows one to write down the conservation laws.
The eigenvalues of the monodromy matrix of solutions $\Psi$ are the gauge invariant.
Represent solutions of (1.\ref{ls1}) as the P-exponent
\beq{sol}
\Psi(\te,z)=\clR(\te,z)P\exp\left(\frac{\imath}{\nu(z)}\int_0^{\te} L(\theta',z) d\te'\right)~~(x=-te^{\imath\te})\,,
\eq
where  $\clR$ is periodic in $\te$.
The monodromy of $\Psi(\te,z)$ is
$$
\exp\left(\f1{\nu(z)}\int_0^{2\pi}{ L(\theta,z)} d\te\right)\,.
$$
Consider the monodromy in a neighborhood of a  pole $z_a\in\Si$ of $L/\nu$
 with a local coordinate $w_a=z-z_a$.
If
$$
\f1{\nu(w_a)}L(\te,w_a)=\left(\frac{L}\nu\right)^a_{-1}w_a^{-1}+\left(\frac{L}\nu\right)^a_0+
\left(\frac{L}\nu\right)^a_1w_a+\ldots\,.
$$
The Hamiltonains
 \beq{hms}
 H_j^a\sim\tr_V\exp\left(\imath\int_0^{2\pi} \left(\frac{L}\nu\right)^a_j d\te\right)\,.
 \eq
are all in involution. Thus, we have an infinite set of Poisson commuting integrals of
motion.

Let us
 "diagonalize" generic element  $L\to h^{-1}\nu\p h+h^{-1}Lh=\clS$, where ${\mathcal S}$ is an element
of the Cartan subalgebra $\gh\subset\gg$. Then the
solutions of the equation (1.\ref{ls1}) can be represented in the form
\beq{1.28}
\Psi(\te,z)=\clR(\te,z)\exp\left(\frac{\imath}{\nu}\int_0^{\te}{\mathcal S(\theta',z)} d\te'\right)\,.
\eq
%
%
%
Let
$$
\f1{\nu(w_a)}{\mathcal S(\te,w_a)}=\left(\frac{{\mathcal S}(\te)}{\nu}\right)^a_{-1}w_a^{-1}+\left(\frac{{\mathcal S}(\te)}{\nu}\right)^a_0+\left(\frac{{\mathcal S}(\te)}{\nu}\right)^a_1w_a+\ldots\,,
$$
Substitute (\ref{1.28}) into (1.\ref{ls1}).
It follows from (\ref{1.17}), (\ref{1.16}) and (\ref{1.28}) that the diagonal matrix elements of ${\mathcal S}_j^m$ are
the densities of the conservation laws
\beq{hf}
 H^a_{j}\sim\tr_V\exp\left\{\imath\int_{S^1}\left(\frac{{\mathcal S}(\te)}{\nu}\right)^a_j  d\te\right\}\,.
\eq
There is a recurrence
procedure to define the  matrices ${\mathcal S}^a_j$.
Details can be found in \cite{LOZ,MOP}.
%

\subsection{The action}

Consider the $4d$ action on the space
 \beq{4h}
 \clM_{a,j}=T_{a,j}\times W
 \eq
 corresponding to the Hamiltonian system defined above:\footnote{We omit the term $\nu  D\bar k$ since, as we argued above, it is inessential.}
$$
\clS^{AHB}=\f1{2\pi\hbar}\sum_{a,j}\Big(\int_{\clM_{a,j}}(\Phi,D\bA)
-\sum_{a=1}^n\clS^{WZW}(S_a)\de(z_a,\bz_a)
-H_j^a(\Phi)Dt_{a,j}\Big)\,.
$$
Here $H_j^a$ are the Hamiltonians (\ref{1.15}) and $\clS^{WZW}$ is the Wess-Zumino-Witten action
$$
\clS^{WZW}=\oh\int_{S^1} d\te (S, Dgg^{-1})+\frac{c^0}2\left(\int_{S^1} d\te(Dgg^{-1},\p gg^{-1})+
D^{-1}(\p gg^{-1},(Dgg^{-1})^2))\right)\,.
$$
To come to the action on the moduli space of the affine Higgs bundles
$\clH^{aff,\,par}(G )$ (\ref{cfc1}) we need to impose the moment map constraints
(\ref{co11}) and fix the gauge.
To do it one should introduce in the action the terms containing the ghost and the anti-ghost fields. Instead, we first fix the gauge and rewrite the action in terms of the fields $L$ and $\bL$.
The action takes the form
$$
\clS^{AHB}=\f1{2\pi\hbar}\sum_{j,a}\Big(\int_{\clM_j^a}(L,D\bL)
-\sum_{a=1}^n\clS^{WZW}(S_a)\de(z_a,\bz_a)
-\sum_{a,j}H_j^a Dt_{a,j}\Big)\,,
$$
and then we impose the moment constraints (\ref{hb2}).

\section{Examples}
In all examples we consider the trivial $S^1$ bundles and put $\bar\mu=0$.

\subsection{Hamiltonians in ${\rm sl}_2$ case.}
Consider the one marked point case.  Then $c^{(0)}=0$,
Due to (\ref{co12}) $\partial_{\bz}\nu=0$ and, therefore, $\nu(z,\bz)=const=\nu_0$.

Let us perform the gauge transformation
\beq{b1}
f^{-1}L f+\nu_0 f^{-1}\p_\te f=L'\,,
\eq
with $f$ defined as follows:
\beq{b2}
f=\mat{\displaystyle{ \sqrt{L_{12}} }}{\displaystyle{ 0 }}
{\displaystyle{ -\frac{L_{11}}{\sqrt{L_{12}}}-\nu_0\frac{\p_\te \sqrt{L_{12}}}
{L_{12}} }}{\displaystyle{ \frac{1}{\sqrt{L_{12}}} }}
\,.
\eq
Then the Lax matrix $L$ is transformed into
\beq{b3}
L'=
\left(
\begin{array}{l}
\ 0 \ \ \ \  1
\\
\ T \ \ \ \ 0
\end{array}
\right)\,,
\eq
where
\beq{b4}
T=L_{21}L_{12}+L_{11}^2+\nu_0\frac{L_{11}\p_\te L_{12}}
{L_{12}}-\nu_0\p_\te L_{11}-\frac{1}{2}\nu_0^2
\frac{\p_\te^2L_{12}}{L_{12}}+\frac{3}{4}\nu_0^2
\frac{(\p_\te{L_{12}})^2}{L_{12}^2}\,.
\eq
The linear problem
\beq{b5.1}
\left\{
\begin{array}{l}
(\nu_0\p_\te+L')\psi=0\,,
\\
(\p_j+M_j')\psi=0\,,
\end{array}
\right.
\eq
where $\psi$ is the Bloch wave function $\psi=exp\{-i\oint\chi\}$,
leads to the Riccati equation:
\beq{b7}
i\nu_0\p_\te\chi-\chi^2+T=0.
\eq
The decomposition of $\chi(z)$ provides densities of the conservation laws
(see \cite{DMN}):
\beq{b8}
\chi=\sum\limits_{k=-1}^{\infty}z^k\chi_k\,,
\eq
\beq{b9}
H_k\sim\oint d\te\chi_{k-1}\,.
\eq
The values of $\chi_k$ can be found from (\ref{b7}) using the expression
(\ref{b4}) for
 $T(z)=\sum\limits_{k=-2}^{\infty}z^k T_{k}$ in a neighborhood of zero.
For $k=-2,\ -1$ and $0$ we have:
\beq{b10}
\left\{
\begin{array}{l}
\chi_{-1}=\sqrt{T_{-2}}=\sqrt{h}\,,
\\
2\sqrt{h}\chi_{0}=T_{-1}+i\nu_0\p_\te\chi_{-1}=T_{-1}\,,
\\
2\sqrt{h}\chi_{1}=T_0+i\nu_0\p_\te\chi-\chi_0^2\,.
\end{array}
\right.
\eq
In Subsections 7.2, 7.3 below, explicit formulae for $T_k$ are used for
the computation
of the Hamiltonians for the elliptic 2d Calogero-Moser and the elliptic
 Gaudin models.

\subsection{Landau-Lifshitz equation (LL)}

In this case $G=\SLT$.
 Let $\Si=\Si_\tau=\mC/(\mZ+\tau\mZ)$ be the elliptic curve with one marked point $z=0$.
 Then the orbit has the form
$\clO=\{S=gp^{(0)}g^{-1}\}$, and $c^{(0)}=0$ (\ref{to}), i.e. $S$ is a traceless $2\times 2$ matrix.

Impose the following  quasi-periodic properties (boundary conditions) on the fields.
Here we use the basis of the Pauli matrices
 $\si_a$ $(a=0,\dots,3)$:
$$
 \si_0=\left(
\begin{array}{cc}
    1 & 0 \\
    0 & 1 \\
  \end{array}
\right)\,,~
 \si_1= \left(
  \begin{array}{cc}
    0 & 1 \\
    1 & 0 \\
  \end{array}
\right)\,,~
\si_2 = \left(
  \begin{array}{cc}
    0 & -\imath \\
   \imath & 0 \\
  \end{array}
\right)\,,~
 \si_3= \left(
  \begin{array}{cc}
    1 & 0 \\
    0 & -1 \\
  \end{array}
\right)\,.
$$

By the gauge transformations $f(z,\bz,\te)$ the field $\bA$ can be made $z$-independent.
 Due to the boundary conditions (\ref{qpc1})  $\bL=0$, so that
$$
f(z,\bz,\te)(\p_{\bz}+\bA(z,\bz,\te))f^{-1}(z,\bz,\te)=0\,.
$$
Then the Lax operator of the  LL equation is defined as
 $$
f(z,\bz,\te)(\nu_0\p_{\te}+\Phi(z,\bz,\te))f^{-1}(z,\bz,\te)=\nu_0\p_{\te}+L^{LL}(z,\bz,\te)\,.
$$
It satisfies the moment map equation
$$
\p_{\bz}L^{LL}(z,\bz,\te)=S(\te)\de(z,\bz)
$$
and has the quasi-periodicities as the Higgs field $\Phi$  in Table 2.
\begin{center}
\textbf{Table 2}: Quasi-periodicities of LL fields
\end{center}
\vspace{-0.5cm}
\beq{qpc1}
\begin{tabular}{|c|c|c|c|}
  \hline
    & &$z\to z+1$& $z\to z+\tau$\\
    \hline
  1&$\bA$ & $\Ad_{\si_3}\bA(z,\bz,\te)$ &$ \Ad_{\si_1}\bA(z,\bz,\te)$ \\
 2& $\Phi$ & $\Ad_{\si_3}\Phi(z,\bz,\te)$ &$ \Ad_{\si_1}\Phi(z,\bz,\te)$ \\
 3& $\ep $&$ \Ad_{\si_3}\ep(z,\bz,\te) $&$\Ad_{\si_1}\ep(z,\bz,\te)$\\
   \hline
\end{tabular}
\eq

To write it down we use the Kronecker elliptic function related to the curve $\Si_\tau$:
\beq{phi}
 \phi(u,z)=\frac{\vth(u+z)\vth'(0)}{\vth(u)\vth(z)}\,,
 \eq
 where $\vth(z)$ is  the theta-function
\beq{theta}
\vth(z|\tau)=q^{\frac
{1}{8}}\sum_{n\in {\bf Z}}(-1)^ne^{\pi i(n(n+1)\tau+2nz)}\,,\quad q=\exp\,2\pi\imath\tau\,.
\eq
The Kronecker function has the following  quasi-periodicities:
\beq{A.14}
\phi(u,z+1)=\phi(u,z)\,,~~~\phi(u,z+\tau)=e^{-2\pi \imath u}\phi(u,z)\,,
\eq
and has the first order pole at $z=0$
\beq{rk}
\displaystyle{
\phi(u,z)=\f1{z}+\frac{\vth'(u)}{\vth(u)}+O(z)\,.
}
\eq
It is related to the Weierstrass function $\wp$ as follows:
\beq{wpphi}
\phi(u,z)\phi(-u,z)=\wp(z)-\wp(u)\,.
\eq
Let
$$
\varphi_1(z)=\phi(\oh,z)\,,\quad\varphi_2(z)=\exp(\pi\imath z)\phi(\frac{1+\tau}2,z)\,,\quad
\varphi_3(z)=\exp(\pi\imath z)\phi(\frac{\tau}2,z)\,.
$$
The Lax operator assumes the form
\beq{sll}
L^{LL}(z,\bz,\te)=\sum\limits_{\al=1}^3 L_\al(z,\te)\si_\al\,,\quad L_\al(z,\te)=S_\al(\te)\varphi_\al(z)\,.
\eq
%
%
The symplectic form $\Om$ (\ref{psf}) is reduced ro the symplectic form on the orbit $\clO(p^{(0)},0)$
(\ref{1.6a}):
\beq{sfl}
\ti\Om=\om(p^{(0)},0)=-\int_{S^1} D(S(p^{(0)},0) g^{-1}D g )
\eq
The Hamiltonian $H^{LL}_2$ (\ref{b9}) assumes the form
$$
H_2=\f1{2}\int_{S^1}d\te\sum_{\al}\Bigl(S_\al(\te)\wp_\al S_\al(\te)+
\bigr(\frac{\nu_0}{2p^{(0)}}\p_\te S_\al(\te)\bigr)^2\Bigr)\,,
$$
where
$\wp_\al$ are the values of the Weierstrass functions at the half-periods.
It is the Hamiltonian of the Euler-Arnold top on the group $L(G )$ defined by the inverse inertia tensor
$$
J=\sum_\al(-(\frac{\nu_0}{2p^{(0)}}\p_\te)^2 +\wp_\al)\,:\,L^*(G )\to L(G )\,.
$$
The corresponding equations of motion (see (\ref{pb})) are the Landau-Lifshitz equations:
\beq{eLL}
\p_t S=[ S,J(S)]+[S,(\frac{\nu_0}{2p^{(0)}}\p_\te)^2 S]
\,.
\eq


\subsection{Calogero-Moser field theory (CM)}

Again, consider the one point case on the elliptic curve $\Si_\tau$ and the trivial $\widehat{\rm SL}(2,\mC)$ bundle
over $\Si_\tau$. It has a moduli space $Bun_{\SLT}\sim\mC/\mZ+\tau\mZ$.
Let $u=u(\theta)$ be a coordinate on the moduli space $Bun_{\SLT}$, and denote $\bfe(u)=\exp\,2\pi\imath u\si_3$.
Assume that the fields have
 the following quasi-periodicities:
\begin{center}
\textbf{Table 3}: Quasi-periodicities of CM fields
\end{center}
\vspace{-0.5cm}
\beq{qpc2}
\begin{tabular}{|c|c|c|c|}
  \hline
    & &$z\to z+1$& $z\to z+\tau$\\
    \hline
  1&$\bA(z,\bz,\te)$ & $\bA(z,\bz,\te)$ &$ \Ad_{\bfe(-u)}\bA(z,\bz,\te)$ \\
 2& $\Phi(z,\bz,\te)$ & $\Phi(z,\bz,\te)$ &$ \Ad_{\bfe(-u)}\Phi(z,\bz,\te)$ \\
 3& $\ep(z,\bz,\te) $&$ \ep(z,\bz,\te) $&$\Ad_{\bfe(-u)}\ep(z,\bz,\te)$\\
   \hline
\end{tabular}
\eq
 

\noindent For stable bundles the orbits of the gauge transformations (\ref{fg}) $\bA\stackrel{f}\to\bL$ are
parameterized  by the $z$-independent diagonal matrices $\bL$. Let us take them in form
\beq{bl}
\bL=\frac{2\pi\imath}{\tau-\bar{\tau}}\,\di(u,-u)\,.
\eq
%
%
As above, we have $\nu=\nu_{0}$.
The solution of the moment map equation (\ref{hb2})
$$
 \p_{\bz}L-\nu_{0}\p_\te\bL+[\bL,L]=\de(z,\bz)S \,,
$$
 where $L=f^{-1}\nu_{0}\p f+f^{-1}\Phi f$ is the Lax operator.
 We should the factorised solutions of this equation with respect to the action of the residual gauge group that preserves the gauge fixing (\ref{bl}).
 It is the group constant diagonal matrices $\clG^{res}=\clH$ - the Cartan subgroup of $\SLT$.
 It acts on the symplectic form (\ref{psf})
 $$
 \f1{\pi}\int_{S^1}\left(\int_{\Si_\tau}(DL,D\bL)- D(S(p^{(0)},0) g^{-1}D g )\right)
 $$
 producing the moment map constraint
 $$
 S_3-2\pi\imath u_\te=0\,,\quad u_\theta=\p_\theta u\,.
 $$
 In addition the gauge fixing of the $\clG^{res}$ action allows one to choose
 $S^+=S^-=l(\te)$. Then
 $$
 S=\left(
     \begin{array}{cc}
       u_\te & l \\
       l & -u_\te \\
     \end{array}
   \right)\,.
 $$
  Then the solution of the moment equation assumes the form
\beq{5.7}
L^{CM}=
\left(
\begin{array}{l}
-\frac{1}{4\pi\imath}v-u_\te E_{1}(z)\ \ \ \ \ \ \ l\phi(2u,z)
\\
\ \ \
l\phi(-2u,z)\ \ \ \ \ \ \ \frac{1}{4\pi\imath}v+u_\te E_{1}(z)
\end{array}
\right)\,,
\eq
where $E_{1}(z)=\p_z\vartheta(z)/\vartheta(z)$ is the first Eisenstein function.

The Hamiltonian of the elliptic Calogero-Moser (ECM) field theory is the integrable 2d continuation of the standard
two-particle ECM Hamiltonian (a motion of particle in the Lame potential)
\beq{clh}
H=-\frac{v^2}{16\pi^2}-l^2\wp(2u)\,,~~(\{v,u\}=1)\,,
\eq
where $\wp(2u)$ is the Weierstrass function. In the field case we have the canonical Poisson bracket $\{v(\theta),u(\theta')\}=\delta(\theta-\theta')$.
From (\ref{b9}) and (\ref{b10}) one finds
\beq{b11}
H_0^{CM}=\int_{S^1}d\te2\sqrt{h}\chi_{1}=\int_{S^1}d\te
(T_0-\frac{1}{4h}T_{-1}^2)=
\eq
$$
=\int_{S^1}d\te\left(-\frac{v^2}{16\pi^2}(1-\frac{u_\te^2}{h})
+(3u_\te^2-h)\wp(2u)-\frac{u_{\te\te}^2}{4l^2}\right)\,,
$$
where $h=u^2_\te+l^2$. For $v$ and $u$ it is the Hamiltonian (\ref{clh}).
The equations of motion produced by $H_0^{CM}$ are of the form:
\beq{b13}
\left\{
\begin{array}{l}
u_t=-\frac{v}{8\pi^2}(1-\frac{u_\te^2}{h})\,,
\\ \ \\
v_t=\frac{1}{8\pi^2 h}\p_\te(v^2u_\te)-2(3u_\te^2-h)\wp'(2u)+
6\p_\te(u_\te\wp(2u))+
\frac{1}{2}\p_\te(\frac{u_{\te\te\te}l-l_\te u_{\te\te}}{l^3})\,.
\end{array}
\right.
\eq
There exists a transformation $\Xi$ of the Lax operators\,:
$$
\Xi\circ(\nu_0\p_\te+ L^{CM})=(\nu_0\p_\te +L^{LL})\circ\Xi\,,
$$
such that solutions of (\ref{b13}) become solutions of the LL equation. $\,(u,v)\to (S_\al,~\al=1,2,3)$
\cite{AZ}. It was called the symplectic Hecke correspondence for integrable systems \cite{LOZ} and can be described in terms of solutions of the extended Bogomolny equation \cite{KW,LOZ1}.
In the 2d case one should define the affine version of the extended Bogomolny equation. We will come to this point in a separate publication.


\subsection{Gaudin field theory and principal chiral model}
%
\def\sln{{\rm sl}(N, {\mathbb C})}
\def\res{\mathop{\hbox{Res}}\limits}

The Gaudin models in classical mechanics are described by the Higgs fields (i.e. the Lax matrices) with a set of simple poles at punctures on a base curve with local coordinate $z$. For elliptic models the latter is the elliptic  curve $\Sigma_\tau$ with punctures $z_a$. Then the Lax matrix is fixed by
a chose of coadjoint orbits $$S^a=\res\limits_{z=z_a}L(z)$$ attached to punctures together with some  boundary conditions (or the quasi-periodic behaviour). See \cite{TZ} for a review of models related to ${\rm SL}$-bundles and \cite{LOSZ} for a generic complex Lie group $G$. Similarly, in the 1+1 field case the Gaudin type models are generalizations of the previously given examples to a multi-pole Higgs field.

\paragraph{Principal chiral model.} The rational 2d field Gaudin model corresponding to Riemann sphere with two punctures is the widely-known principal chiral model. Indeed, consider the Zakharov-Shabat equation\footnote{In this subsection we put $\nu_0=1$ for simplicity.}
\beq{a1}
\p_{t}L(z)-\p_\te  M(z)=[L(z),M(z)]\,,
\eq
with
\beq{a2}
L(z)=\frac{S^1}{z-z_1}+\frac{S^2}{z-z_2}\,,\qquad M(z)=\frac{S^1}{z-z_1}-\frac{S^2}{z-z_2}\,.
\eq
The we have equations of motion
\beq{a3}
\left\{\begin{array}{l} \p_t S^1-\p_\te  S^1=-\frac{2}{z_1-z_2}[S^1,S^2]\,,\\
\p_t S^2+\p_\te  S^2=\frac{2}{z_1-z_2}[S^1,S^2]\,,
\end{array}\right.
\eq
which are generated by the Poisson brackets
\beq{a31}
\{S^a_\al(x),S^b_\be(y)\}=2\sqrt{-1}\delta^{ab}\ve_{\al\be\ga}S^a_\ga(x)\delta(x-y)
\eq
 and the Hamiltonian
\beq{a4}
H=\int\limits_{{\mathbb
S}^1}d\te\ \left(P_1-P_2-\frac{\langle S^1
S^2\rangle}{z_1-z_2}\right)\,.
\eq
Here $\int_{S^1}d\te\
P_a$ is the shift operator in the loop algebra ${\hat{\rm sl}(N,
{\mathbb C})}$:
\beq{a41}
\{\int_{S^1}d\te'\
P_a(\te'),S^b(\te)\}=\delta_{ab}\p_\te S^b(\te)\,.
\eq
 The substitution $S^1=\frac{1}{2}(l_0+l_1)$ and $S^2=\frac{1}{2}(l_0-l_1)$
transforms (\ref{a3}) into equation of the principal chiral model:
\beq{a5} \left\{\begin{array}{l} \p_tl_1-\p_\te  l_0+\frac{2}{z_1-z_2}[l_1,l_0]=0\,,\\
\p_tl_0-\p_\te l_1=0\,.
\end{array}\right. \eq
Also, by changing the coordinates  $(\theta,t)$ to the "light-cone" coordinates
$\xi=\frac{t+\theta}{2},\ \eta=\frac{t-\theta}{2}$, one gets
\beq{a6} \left\{\begin{array}{l} \p_\eta S^1=-\frac{2}{z_1-z_2}[S^1,S^2],\\
\p_\xi S^2=\frac{2}{z_1-z_2}[S^1,S^2].
\end{array}\right. \eq

\paragraph{Elliptic 1+1 Gaudin model: first flows.} Let us proceed to the elliptic case. The multi-pole extensions of the (spin) Calogero-Moser field theory were studied in \cite{LOZ}. Here we briefly review results of \cite{Z} on the multi-pole generalization of ${\hat{\rm sl}(2,
{\mathbb C})}$-valued Lax matrix (\ref{sll}) with the quasi-periodic properties (\ref{qpc1}):
\beq{a61}
L(z)=\sum\limits_{c=1}^n\sum\limits_{\ga=1}^3\si_\ga
S^c_\ga \vf_\ga(z-z_c)\,.
 \eq
Using (\ref{b7})-(\ref{b9}) one gets the following ''first flow'' Hamiltonians:
\beq{a7}
 H_{a,1}= \oint\limits_{{\mathbb
S}^1}d\te\ (P_a+H_a)\,,
\eq
\beq{a71}
H_a=-\frac{1}{2}\sum\limits_{c\neq a}\langle
S^a\hat\vf_{ac}(S^c)\rangle=-\sum\limits_{c\neq a} S^a_1S^c_1\vf_1(z_a-z_c)
+S^a_2S^c_2\vf_2(z_a-z_c)+S^a_3S^c_3\vf_3(z_a-z_c)\,.
\eq
Here and below we use the following notations for the linear operators:
\beq{a8}
\hat\wp:\ S_\al\rightarrow S_\al\wp(\om_\al),\ \ \
\hat\vf_{ab}:\ S_\al\rightarrow S_\al\vf_\al(z_a-z_b),\ \ \ \hat
F_{ab}:\ S_\al\rightarrow S_\al F_\al(z_a-z_b)\,,
\eq
where $F_\al(z)=\vf_\al(z)(E_1(z)+E_1(\om_\al)-E_1(z+\om_\al))$.

The Hamiltonians (\ref{a71}) generate dynamics described by the following equations:
\beq{a9}\left\{
\begin{array}{l}
\p_{{t}_a}S^a-\p_\te  S^a=-\sum\limits_{c\neq
a}[S^a,\hat{\vf}_{ac}(S^c)],
\\
\p_{{t}_a}S^b=[S^b,\hat{\vf}_{ba}(S^a)].
\end{array}
\right.
\eq
These equations are equivalent to Zakharov-Shabat equation (\ref{a1}) with $L(z)$ (\ref{a61}) and
\beq{a10}
 M_a(z)=\sum\limits_{\ga=1}^3\si_\ga S^a_\ga
\vf_\ga(z-z_a)\,. \eq

\paragraph{Elliptic version of the Principal chiral model.} Consider the case of two punctures (i.e. $n=2$). Then
$L(z)=M_1(z)+M_2(z)$. Let us choose $M(z)=M_1(z)-M_2(z)$. The above equations yield (with $\p_t=\p_{t_1}-\p_{t_2}$)
\beq{a11} \left\{\begin{array}{l} \p_t S^1-k\p_\te  S^1=-2[S^1,\hat\vf_{12}(S^2)],\\
\p_t S^2+k\p_\te  S^2=2[S^2,\hat\vf_{21}(S^1)].
\end{array}\right. \eq
 or by analogy with (\ref{a6}):
\beq{q06008} \left\{\begin{array}{l} \p_\eta S^1=-2[S^1,\hat\vf_{12}(S^2)],\\
\p_\xi S^2=2[S^2,\hat\vf_{21}(S^1)].
\end{array}\right. \eq

\paragraph{Elliptic 1+1 Gaudin model: second flows (coupled Landau-Lifshitz equations).} The second flows are described by the following set of Hamiltonians:
\beq{a12}
\begin{array}{c}
\displaystyle{
 H_{a,2}=\int_{S^1}d\theta\
\left(\frac{1}{4}\langle
S^a\hat\wp(S^a)\rangle+\frac{1}{2}\sum\limits_{c\neq a}\langle S^a
\hat{F}(S^c)\rangle-\frac{1}{4}\langle\left(\sum\limits_{c\neq
a}\hat\vf_{ac}(S^c)\right)^2\rangle+\right.
}
\\ \ \\
\displaystyle{
\left.+\frac{1}{8\la_a^2}\left(\sum\limits_{c\neq a}\langle
S^a\hat\vf_{ac}(S^c)\rangle\right)^2-
\frac{1}{4\la_a^2}\sum\limits_{c\neq a}\langle
\hat\vf_{ac}(S^c)\p_\te S^aS^a\rangle+\frac{1}{16\la_a^2}\langle\left(\p_\te S^a\right)^2\rangle
\right)\,,
}
\end{array}
\eq
where $\lambda_a$ are the eigenvalues of $S^a$ (i.e. spectrum of $S^a$ is ${\rm diag}(\lambda_a,-\lambda_a)$), and it is
assumed that $\p_\te \lambda_a=0$. Equation of motion take the form
\beq{a13} \left\{
\begin{array}{l}
\displaystyle{
\p_{{\tilde t}_a} S^a-\p_\te
\eta^a=[S^a,\hat{\wp}(S^a)]+\sum\limits_{c\neq
a}[\eta^a,\hat{\vf}_{ca}(S^c)]-\hat{\vf}_{ca}([S^c,\hat{\vf}_{ca}(S^a)]),
}
\\
\displaystyle{
\p_{{\tilde t}_a}
S^b=[\hat{\vf}_{ab}(\eta^a),S^b]+\hat{\vf}_{ba}([\hat{\vf}_{ba}(S^b),S^a])\,,
}
\end{array}\right.
\eq
where
\beq{a14}
\displaystyle{
 \eta^a=-\frac{1}{4\la_a^2}[S^a,\p_\te S^a]+\sum\limits_{c\neq
a}\hat{\vf}_{ac}(S^c)+\frac{H_a}{\la_a^2}S^a\,.
}
\eq
In the case of a single marked point ($n=1$) we get the Landau-Lifshitz equation in the form:
\beq{a15}
\displaystyle{
 \p_t S+\frac{1}{4\la^2}[S,S_{\te\te}]=[S,\hat{\wp}(S)]\,,
 }
  \eq
described by the Hamiltonian
\beq{a16}
\displaystyle{
H=\oint\limits_{{\mathbb S}^1}d\te\
\left(\frac{1}{4}\langle
S\hat\wp(S)\rangle+\frac{1}{16\la^2}\langle\left(\p_\te S\right)^2\rangle\right).
}
\eq
Similarly, one can write down in the trigonometric and rational cases. For example, in the straightforward
 rational limit (related to XXX 6-vertex $R$-matrix) the above equations provide
 the model of coupled Heisenberg magnets. The rational 11-vertex deformation was described in \cite{LOZ14}.
 Trigonometric 6-vertex and 7-vertex models are described in the same way.



\section{Correspondence between 4d-CS and AHB}


Consider expansion (\ref{1.16}) of the Hamiltonian $\clH(L)$ (\ref{1.17}):
$$
H(L)=\sum_{a}\sum_{j=-1}^{+\infty}H^a_j(L)w_a^j\,.
$$
 Let us pass to the following new field:
\beq{blp}
\bL'_{a,j}=\bL-\frac{\de H^a_j(L)}{\de L}t_{a,j}\,.
\eq
Since $\bL$ satisfies 2.(\ref{cc}) then $\bL'_{a,j}$ satisfies the equation
\beq{lp}
\bp M_{a,j}-\p_{a,j}\bL'_{a,j}+[M_{a,j},\bL'_{a,j}]=0\,,~~
\left([D_{\bL'},D_{M_{a,j}}] =0\right)\,.
\eq
To prove it we use the equation
$$
\p_{a,j}\frac{\de H^a_j(L)}{\de L}+[M_{a,j},\frac{\de H^a_j(L)}{\de L}]=0\,.
$$
The latter follows from (\ref{1.20})  and from (\ref{lf}).
%

Consider a family of 3d spaces with coordinates
 \beq{w}
 \clW_{a,j}=\{(\bz,T_{a,j},\te\in S^1)\}\subset\clM_{a,j}~(\ref{4h})
 \eq
 and  the $\clP$-bundle over $\clW_{a,j}$ with connections
 \beq{co}
 D_{\clA_{a,j}}=(D_{A_\te}d\te\,,\, D_{M_{a,j}}dt_{a,j}\,,\,D_{\bL'_{a,j}}d\bz)\,,
 \eq
 $$
 D_{A_\te}=\p_\te+A_\te\,,\quad D_{M_{a,j}}=\p_{a,j}+M_{a,j}\,, \quad D_{\bL'_{a,j}}=\bp+\p_\te +\bL'_{a,j}\,.
 $$

It follows from (\ref{fg2}) that the system (\ref{cc}) assumes the form:
\beq{CSE}
\begin{array}{ll}
        1.& \nu[D_{M_{a,j}},D_{A_\te}] =0\,,  \\
2.& [D_{\bL'_{a,j}},D_{M_{a,j}}] =0
\,, \\
3. &  \nu[D_{\bL'_{a,j}},D_{A_\te}]=\sum_{a=1}^n\de(z-z_a)S_a \,.
        \end{array}
\eq
The delta-functions in the r.h.s of (3.\ref{CSE}) mean that the connection form (i.e. $L$) has the first order
poles.
Equations (\ref{CSE}) are the equations of motion for the 4d-CS action on the 4d spaces
 $\clM_{a,j}$ (\ref{4h})
  $$
  S_{4d} =\f1{2\pi\hbar}\int_{\clM_{a,j}}\nu\cdot CS(\clA_{a,j})\,,
  $$
  where $\clA_{a,j}=(D_{M_{a,j}},D_{\bL'_{a,j}},D_{A_\te})$  and
$CS(\clA_{a,j}) := \tr\Bigl( \clA_{a,j} \wedge d\clA_{a,j} +
\frac{2}{3} \clA_{a,j} \wedge  \clA_{a,j} \wedge  \clA_{a,j}\Bigr)$.
Thereby, we rewrite the equations  (\ref{cc}) of the AHB theory in the Chern-Simons form (\ref{cs4}).

Comparing the system (\ref{CSE}) with the system (\ref{ecs}) in 4d-CS theory we come to the following relations between the fields in these two constructions:
\begin{center}
\textbf{Table 4}: Correspondence between fields
\end{center}
\vspace{-0.5cm}
\beq{nc}
\begin{tabular}{|c|c|}
  \hline
4d\, CS & AHB\\
 \hline
 $ M=R^2\times\Si$ & \,$\clM_{a,j}\,~~$(\ref{4h})\\
  $(w,\bar w)\times(z,\bz) $\,& $(t_{a,j},\te)\times(z,\bz)$\\
  $\bA=0$ & $\bL'_{a,j}$~~(\ref{blp})
   \\
  $A_w\,,\,   A_{\bar w}$
  &
 $A_\te\,,\,  M_{a,j}$
  \\
 $ \om$ & $\nu\,~~$(\ref{nu})  \\
  $\phi_a $& $S_a\in\clO_a$\\
   \hline
\end{tabular}
\eq

Thus, we established the equivalence of two constructions at the classical level in the case
when the surface defects correspond to the first order poles and the  $W$ bundles (\ref{wb}) are trivial.
%

\section{Appendix}


\subsection{Affine Lie algebras \cite{Ka}}

\setcounter{equation}{0}
\def\theequation{A.\arabic{equation}}

Let $\gg$ be a simple complex Lie algebra and $L(\gg)=\gg\otimes\mC(x)$, $x\in \mC^*$
 is the loop algebra of Laurent polynomials. Let $(\,,\,)$ be
 an invariant form on $\gg$. And let res be the coefficient $c_{-1}$ in the Laurent expansion of $X=\sum c_kx^k\in L(\gg)$. Define the form on $L(\gg)$
  $$
  \lan X\,,\, Y\ran=\int_{S^1}(X,Y)d\te\,.
  $$
Consider its central extension  $\hat{L}(\gg)=\{(X(x),k)\}$, $k\in\mC$.
 The commutator in  $\hat{L}(\gg)$ assumes the form
 $$
[(X_1,k_1),(X_2,k_2)]=([(X_1,X_2)]_0,\lan X_1,\p X_2\ran)\,,~~(\p=\imath x\p_x )\,,
 $$
where $[(X_1,X_2)]_0$ is a commutator in $\gg$,

 The  cocentral
extension $\check{L}(\gg)$ of $\hat L(\gg)$ is the algebra
 \beq{lcc}
\check{L}(\gg)=\{\clX=(X\,,k\,,\mu)=(\mu\p+X,k)\,,~X\in L(\gg)\,,~~k\in\mC\,,~\mu\in\mC\}\,.
 \eq
The commutator in $\check{L}$ assumes the form
 \beq{lc2}
[\clX_1,\clX_2]=[(X_1,k_1,\mu_1),(X_2,k_2,\mu_2)]= (\mu_1\p X_2-\mu_2\p X_1+[X_1,X_2]_0\,,
\lan X_1,\p X_2\ran\,,0 )\,.
 \eq
There is invariant non-degenerate form on $\check{L}$
 \beq{inm}
(\clX_1,\clX_2)=\lan X_1,X_2\ran+k_1\mu_2+k_2\mu_1\,.
 \eq
 Let $K$ be a generator of the central charge and $\gh^0$ is the Cartan subalgebra
 of $\gg$. The Cartan subalgebra $\gh$ of $\check{L}$ takes the form
 \beq{cs}
\gh=\gh^0\oplus \mC\p\oplus\mC K\,.
 \eq
%

\vspace{2mm}

Let $L(G )$ be the loop group corresponding to the loop Lie algebra $L(\gg)$
  \beq{logr}
 L(G )=G \otimes \mC(t))=\left\{\,\sum_kg_kx^k\,,~g_k\in G \,\right\}\,,
  \eq
  The central extension $\hat L(G )=\{g(x),\zeta\}$ is defined by the
  with the multiplication
\beq{ml}
(g,\zeta)\times(g',\zeta')=\left(gg',\zeta\zeta' {\mathcal C}(g,g')\right),
\eq
where ${\mathcal C}(g,g')$ is a 2-cocycle on $L(G )$
 providing the associativity of the multiplication.

Consider   the shift operators $T_\mu=\exp(\mu\p)$, $\mu\in\mC$ acting on $L(G )$. The semidirect product is
the  co-central extension of $\hat L(G )$
\beq{cce}
 \check L(G )=\hat L(G )\rtimes\{T_\mu\}\,.
 \eq
%
%
The adjoint action of $f\in L(G )$ is defined as
\beq{aa}
\Ad_f\clX=\Ad_f(X,k,\mu)=(f Xf^{-1}-\mu\p f f^{-1}\,,\,k+\lan f^{-1}\p f,X\ran-\oh\mu\lan(f^{-1}\p f)^2\ran\,,\mu)\,.
\eq
The coalgebra
\beq{lc}
\check{L}^*(\gg)=\{\clY=(Y,r,\nu)\sim(\nu\p+Y,r)\}
\eq
 is defined by the pairing
 \beq{pa}
(\clX,\clY)=\lan X,Y\ran +k\nu+\mu r\,.
\eq
Here $Y$ is a one form $Yd\te$ on $S^1$.

The coadjoint  action of $L(G )$
 assumes the form
\beq{ca}
\Ad^*_f\clY=\Ad^*_f(Y,r,\nu)=
(f^{-1}Y f+\nu f^{-1}\p f\,,\,r-\lan\p f f^{-1},Y\ran-\oh\nu\lan(f^{-1}\p f)^2\ran\,,\,\nu)\,.
\eq
The corresponding Lie algebra $L(\gg)\otimes \mC[x,x^{-1}]\{\ep\}$ acts as
\beq{aaa}
\ad_\ep\clX=( [\ep, X]_0-\mu\p \ep,k+\lan\p \ep, X\ran,0)\,.
\eq
\beq{aca}
\ad^*_\ep\clY=
([Y,\ep]_0+\nu\p \ep,r-\lan\p \ep, Y\ran,0)\,.
\eq


\subsection{Coadjoint orbits}

Coadjoint orbits are results of coadjoint action (\ref{ca}) of $L(G )$ on a fixed element
$$
\clY^{(0)}=(c^{(0)}\p+p^{(0)},0)=(p^{(0)},0,c^{(0)})
$$
of the Lie coalgebra $\hat L^*(\gg)$ (\ref{lc}).

Consider the orbit of the loop group orbit passing through $\clY^{(0)}$
$$
\Ad^*_g\clY^{(0)}=\left( {\mathcal O}(p^{(0)},c^{(0)})\,,\,-\lan\p gg^{-1}p^{(0)}\ran -\oh c^{(0)}\lan(g^{-1}\p g)^2\ran\,,\,c^{(0)}\right)\,,
$$
where
\beq{c11}
 {\mathcal O}(p^{(0)},c^{(0)})=\{
 S=g^{-1}p^{(0)}g+c^{(0)}g^{-1}\p g\,,~g\in L(G )\}\,,
\eq
%
The symplectic form on the orbit is the Kirillov-Kostant form
\beq{1.6a}
\om^{KK}=-\int_{S^1} (p^{(0)},D g g^{-1}D g g^{-1})+
+\frac{c^{(0)}}{2}\int_{S^1} \left(D g g^{-1},  \p( D g g^{-1})\right)=
\eq
$$
\int_{S^1} (S(p^{(0)},c^{(0)}), g^{-1}D g g^{-1}D g)\,.
$$
The corresponding Poisson brackets are
\beq{pb}
\{S_\al(x),S_\be(y)\}=\de(x/y)c^\ga_{\al\be}S_\ga(x)+c^{(0)}\ka_{\al\be}\p\de(x/y)\,,
\eq
where $\ka_{\al\be}$ is invariant form on $\gg$.
The form $\om^{KK}$ is invariant under transformations
\beq{kki}
g\to gf\,,~f\in L(G )\,.
\eq
 The corresponding moment is  $S(p^{(0)},c^{(0)})$.
The action the $\{\exp(\varepsilon_2(z,\bz)\p\})$ component takes the form (\ref{hvf1})
$$
\de_{\varepsilon_{2}} g=\varepsilon_{2}\p g\,.
$$
The central element $\{\exp (\varepsilon_3)\}$ (\ref{gg1}) does not act on $\clY^{(0)}$.

We assume that $p^{(0)}$ is a semi-simple element in the Cartan subalgebra
$\gh^\mC\subset\gg$. Its centralizer is the Cartan subgroup $H^\mC$.
The invariants defining the orbit ${\mathcal O}(p^{(0)},c^{(0)})$ are the conjugacy classes of the monodromy operator corresponding to the connection $c^{(0)}\p+S$ along a contour in $\mC^*$.
In fact, there
is a one-to-one correspondence between the set of $L(G )$-orbits
and the set of conjugacy classes in the group $G $.
The orbit is the coset space ${\mathcal O}(p^{(0)},c^{(0)})\sim L(G )/H^\mC$
 for $c^{(0)}\neq 0$,
 and ${\mathcal O}(p^{(0)},0)\sim L(G )/L(H^\mC)$, where $H^\mC$ is the Cartan subgroup of $G $.


\subsection*{Acknowledgments}

The work of M. Olshanetsky was partially supported by Russian Science Foundation
grant 21-12-00400.


\noindent

\end{document}